\documentclass[twocolumn,showpacs,preprintnumbers,amsmath,amssymb,prl,superscriptaddress]{revtex4-1}
\usepackage{graphicx}
\usepackage{dcolumn}
\usepackage{bm}
\usepackage{epstopdf}
\usepackage{float}
\usepackage{xcolor}
\usepackage{amsmath}

\begin{document}

\title{Phonon self-energy and origin of anomalous neutron scattering spectra in SnTe and PbTe thermoelectrics}

\author{C.W. Li}
\email{lic2@ornl.gov}
\affiliation{Materials Science and Technology Division, Oak Ridge National Laboratory, Oak Ridge, Tennessee, 37831}
\author{O. Hellman}
\affiliation{Department of Physics, Chemistry and Biology, Link\"oping University, Link\"oping, Sweden}
\author{J. Ma}
\affiliation{Quantum Condensed Matter Division, Oak Ridge National Laboratory, Oak Ridge, Tennessee, 37831}
\author{A.F. May}
\affiliation{Materials Science and Technology Division, Oak Ridge National Laboratory, Oak Ridge, Tennessee, 37831}
\author{H.B. Cao}
\affiliation{Quantum Condensed Matter Division, Oak Ridge National Laboratory, Oak Ridge, Tennessee, 37831}
\author{X. Chen}
\affiliation{Materials Science and Technology Division, Oak Ridge National Laboratory, Oak Ridge, Tennessee, 37831}
\author{A.D.~Christianson}
\author{G. Ehlers}
\affiliation{Quantum Condensed Matter Division, Oak Ridge National Laboratory, Oak Ridge, Tennessee, 37831}
\author{D.J. Singh}
\author{B.C. Sales}
\author{O. Delaire}
\affiliation{Materials Science and Technology Division, Oak Ridge National Laboratory, Oak Ridge, Tennessee, 37831}

\date{\today}

\begin{abstract}
The anharmonic lattice dynamics of rock-salt thermoelectric compounds SnTe and PbTe are investigated with inelastic neutron scattering (INS) and first-principles calculations. The experiments show that, surprisingly, although SnTe is closer to the ferroelectric instability, phonon spectra in PbTe exhibit a more anharmonic character. This behavior is reproduced in first-principles calculations of the temperature-dependent phonon self-energy. Our simulations reveal how the nesting of phonon dispersions induces prominent features in the self-energy, which account for the measured INS spectra and their temperature dependence. We establish that the phase-space for three-phonon scattering processes, rather than just the proximity to the lattice instability, is the mechanism determining the complex spectrum of the transverse-optical ferroelectric mode.
\end{abstract}


\maketitle

Understanding phonon anharmonicity is both of fundamental importance and of practical interest, for instance to improve the efficiency thermoelectrics materials, by suppressing the lattice thermal conductivity, $\kappa_{\rm lat}$. The rock-salt semiconductor compounds PbTe and SnTe are currently among the most efficient thermoelectric materials, in part because of their very low phonon thermal conductivity.\cite{Snyder:2008jh,Pei:2011kx,Biswas:2012fw, Parker:2013js,Xu:2011dt,Akhmedova:2009je, Zhang:2013ca} Several factors have been invoked in explaining this low $\kappa_{\rm lat}$, including soft bonds, heavy masses, as well as proximity to a ferroelectric lattice instability, which causes strong anharmonicity.\cite{Delaire:2011gr,Jensen:2012dz,An:2008cw, Zhang:2011bp, Shiga:2012fu, Tian:2012fd} 

The ferroelectric instability in group IV tellurides originates from the instability of the half-filled resonant {\it p}--band, with the non-linear Te polarizability leading to a strong anharmonicity of ionic potentials.\cite{Lucovsky:1973, Littlewood:1980-1, Littlewood:1980-2, BussmannHolder:1983, Rabe:1985fc, An:2008cw, Zhang:2011bp} This strong anharmonicity of the cubic paraelectric phase causes the frequency of the low energy zone-center transverse optical (TO) phonon to increase markedly with increasing $T$, in a clear departure from the quasi-harmonic theory of lattice dynamics.\cite{Delaire:2011gr, Alperin:1972ik, Jantsch:1983fp, Bruesch:1982, Srivastava:1990, Fultz:2010} Recent neutron scattering measurements have identified a striking additional signature of anharmonicity in the shape of the TO phonon spectrum of PbTe.\cite{Delaire:2011gr, Jensen:2012dz}

In the temperature range $200 \leqslant T \leqslant 500\,$K, the TO spectrum of PbTe exhibits a broad and split peak at the zone center, clearly departing from a damped harmonic oscillator profile.\cite{Delaire:2011gr,Jensen:2012dz, Lovesey:1984} A similar feature in the TO phonon mode was observed in frequency-dependent reflectivity measurements.\cite{Burkhard:1977} It has been suggested that this anomaly is related to a model of dipolar fluctuations, or to the emergence of an additional localized optical mode, beyond the dispersions of the rock-salt structure.\cite{Zhang:2011bp, Jensen:2012dz, Bozin:2010iw} However, its precise nature and origin remain unknown. In addition, an asymmetry of peaks was reported in the pair-distribution-function of PbTe, and was first interpreted in terms of dipolar atomic off-centerings increasing with temperature.\cite{Bozin:2010iw,Kastbjerg:2013gc} However, this interpretation was later invalidated by ab-initio molecular dynamics simulations, which reproduced the feature and showed it arises from anharmonic atomic vibrations.\cite{Zhang:2011bp} Further measurements of the local-structure of PbTe with x-ray absorption fine structure found no off-centerings.\cite{Bridges:2013,Keiber:2013}

In this Letter, we explain the origin of the complex TO spectral function, by comparing measurements in SnTe and PbTe, and performing systematic first-principles simulations of the temperature-dependent phonon self-energy. Our results establish that, surprisingly, the TO mode in SnTe is less anharmonic, even though SnTe is closer to the ferroelectric instability than PbTe. We explain how these results arise from a resonance in the real part of the phonon self energy ($\it\Delta(\Omega)$), which is more pronounced in PbTe than in SnTe, owing to better dispersion nesting.

\begin{figure*}[t]
\includegraphics[width=2.2\columnwidth]{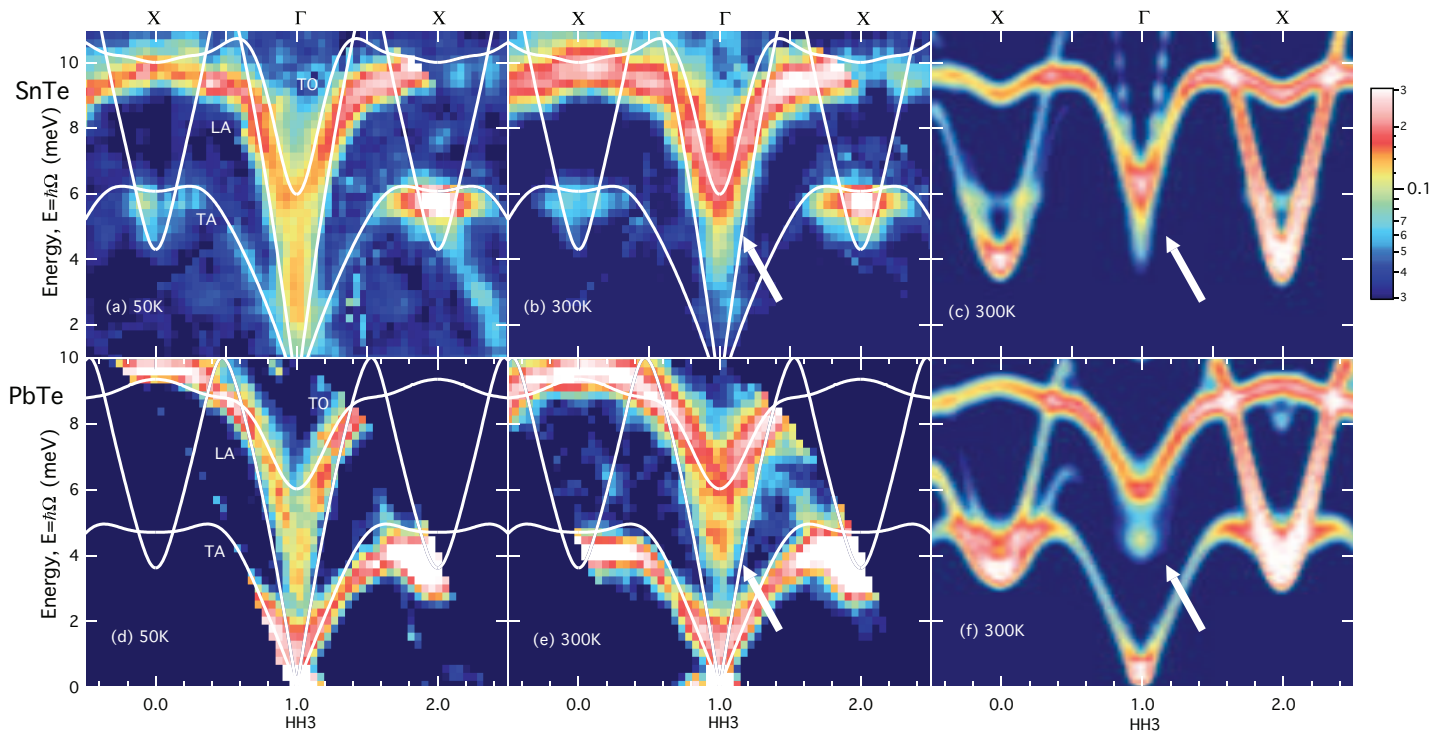}
\caption{\label{fig:SQE_all} $\chi^{\prime \prime}(\mathbf Q, \it\Omega)$ measured with INS and computed with first-principles anharmonic lattice dynamics, showing anomaly of TO mode at the zone center, $\Gamma$ (arrows). (a,b) $\chi^{\prime \prime}(\mathbf Q, \it\Omega)$ of SnTe measured with CNCS, at $T=50$ and $300\,$K, for $\mathbf Q$ along $[H,H,3]$ (intensities are integrated over $\pm0.1\,$\textit{r.l.u.} along $[0,0,1]$ and $[1,\overline{1},0]$ directions, and plotted on a logarithmic-scale). (d,e) Same for PbTe. Part of the data in (e) was shown in Ref. \cite{Delaire:2011gr}. White lines are renormalized phonon dispersions calculated at $T=300\,$K. (c,f) Anharmonic $\chi^{\prime \prime}(\mathbf Q,\it\Omega)$ along $[H,H,3]$, computed from the phonon self-energy (see text).}
\end{figure*}

Anharmonicity renormalizes the spectra of phonon quasiparticles, both shifting the positions of the phonon peaks, as well as introducing finite line widths (phonon damping). The INS spectra are determined by the phonon self-energy $\it\Sigma(\Omega) = \Delta(\Omega) + i \Gamma(\Omega)$, whose real part $\it\Delta$ encodes a shift from the harmonic frequency, while the imaginary part $\Gamma$ encodes the phonon scattering rate, or peak width. The energy transfer between the neutron and phonon system is $E=\hbar \it\Omega$. Anharmonic lattice dynamics simulations for PbTe have shown that the anharmonicity couples the TO mode and the longitudinal acoustic (LA) branch, suppressing the thermal conductivity of PbTe.\cite{An:2008cw, Shiga:2012fu, Tian:2012fd} However these studies did not investigate the full shape of $\it\Sigma(\Omega)$, and the TO splitting remains unexplained. 

Single crystals were grown by a modified Bridgman technique, and characterized with x-ray diffraction and transport measurements (Supplemental Material). The carrier concentration determined from Hall measurements was $n_h \simeq 6.5\pm1.5\times10^{20}\,\rm h/$cm$^{3}$ in SnTe crystals, and $n_e \simeq 1.7 \times 10^{17}\, \rm e/$cm$^{3}$ in PbTe crystals. The resistivity measurement confirmed that the SnTe crystals have a Curie temperature $T_{\rm C} \simeq 42\,$K, consistent with expectations for this carrier density based on prior work, while PbTe remains incipient ferroelectric with an extrapolated $T_{\rm C} \simeq -60\,$K.\cite{Kobayashi:1976, Jantsch:1983fp} 

Inelastic neutron scattering (INS) measurements were performed with the CNCS cold neutron time-of-flight spectrometer at the Spallation Neutron Source and the HB-3 thermal triple-axis spectrometer at the High-Flux Isotope Reactor. CNCS measurements were performed with incident energies $E_{i}=12$ and $25\,$meV, and at two different temperatures (50 and 300$\,$K). The crystals were oriented with the $[\bar{1}10]$ axis vertical, and data were collected for different rotations around this direction, over a wide range of angles. The data were subsequently combined to generate the four-dimensional scattering function, $\it S$$(\mathbf Q,\it\Omega)$, using standard software.\cite{mantid,horace} The imaginary dynamical susceptibility, $\chi^{\prime \prime}(\mathbf Q,\it\Omega) = \it S$$(\mathbf Q,\it\Omega) / (n_T({\it\Omega})+1)$, with $n_T({\it\Omega})$ the Bose distribution, was then ``sliced'' along selected $\mathbf Q$-directions to produce two-dimensional views, shown in Fig.~\ref{fig:SQE_all}. For SnTe, data from both incident energies were combined since lower incident energy provides better resolution, but limited $\mathbf Q$ and $\it\Omega$ coverage. The INS data on this figure are compared with first-principles calculations of the dispersions, discussed below.

\begin{figure}
\includegraphics[width=1.0\columnwidth]{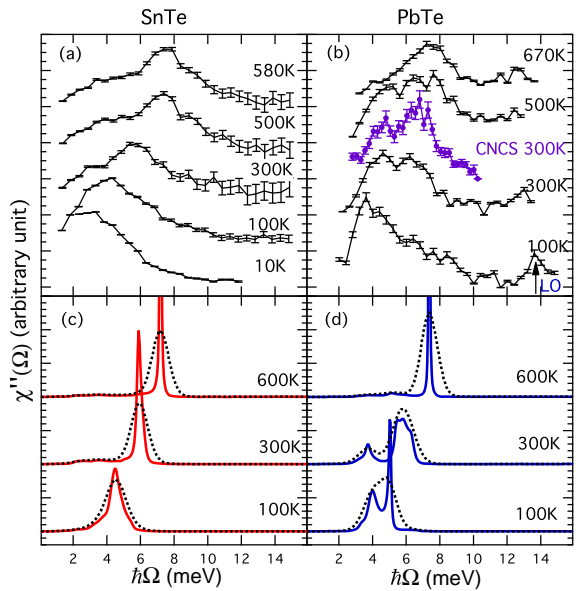}
\caption{\label{fig:TO_113} Evolution of TO phonon spectrum at $\Gamma$--point. (a,b) Phonon spectra of SnTe (a) and PbTe (b) at (1,1,1), measured with HB-3 at a series of temperatures. The purple dots are CNCS measurements at (1,1,3) with intensities integrated over $\pm0.1\,$\textit{r.l.u.} along all $\mathbf Q$ directions. Data in (b) was shown in Ref. \cite{Delaire:2011gr}. (c, d) Phonon spectra (light red: SnTe, dark blue: PbTe) computed from the self-energy, showing the more pronounced split-peak structure in PbTe. Dashed black dash lines are spectra convoluted with instrumental resolution.}
\end{figure}

As can be seen in Fig.~\ref{fig:SQE_all}, the TO phonon dispersion exhibits a pronounced dip at the zone center, ${\Gamma}$, in both PbTe and SnTe, in agreement with the literature.\cite{Daughton:2001hp,Cochran:1966cr,Alperin:1972ik,Cowley:2002ei,Cowley:1996vt,Pawley:1966vs,Strauch:1987ts} It is important to note, though, that our measurements map $\chi^{\prime \prime}(\mathbf Q,\it\Omega)$, instead of just the position of maxima in intensity. These maps reveal a strong broadening for the soft optical phonons at $\mathit{\Gamma}$ in both materials, indicating short phonon lifetimes and large anharmonicity. Because of the larger carrier density in SnTe, no splitting is expected between the longitudinal optical (LO) and (doubly degenerate) TO modes at ${\Gamma}$.\cite{Jantsch:1983fp} This was confirmed by comparing the spectra under different polarization conditions at equivalent $\rm Q$ points near ${\Gamma}$ in several Brillouin zones. On the other hand, the much lower carrier concentration in the PbTe sample leads to the LO--TO splitting at ${\Gamma}$ (see Fig.~\ref{fig:TO_113}-b). Panels Fig.~\ref{fig:SQE_all}-c,f show the results of our anharmonic calculation of $\chi^{\prime \prime}(\mathbf Q,\it\Omega)$ for SnTe and PbTe, respectively (see details below). A long tail is observed at low energy below the optical modes at ${\Gamma}$ with excellent agreement agreement with the experimental results. In PbTe, this low-energy tail splits into a separate maximum at $4\,$meV, while the rest of the TO intensity is at $6\,$meV, continuous with the dispersion as $q \rightarrow\Gamma$. 

In the case of mild anharmonicity, the spectral function of a phonon of wave vector $\mathbf q$ and branch index $j$, $\chi^{\prime \prime}_{\mathbf q j}(\it\Omega)$, presents a single well-defined peak, that approaches a Lorentzian centered at $\hbar \omega_{\mathbf q j} + {\it\Delta}_{\mathbf q j}$, where $\omega_{\mathbf q j}$ is the purely harmonic frequency, and ${\it\Delta}_{\mathbf q j}$ a small anharmonic shift. The full-width-at-half-maximum of the Lorentzian is then inversely related to the phonon scattering rate, $2{\it\Gamma}_{\mathbf q j} = \tau_{\mathbf q j}^{-1}$. However, in both SnTe and PbTe, the TO spectral function deviates strongly from this Lorentzian behavior near the zone center. In particular, the spectrum in PbTe exhibits a double peak in a range of temperature. Such features have been interpreted as a possible ``new mode'' \cite{Jensen:2012dz, Burkhard:1977}, but their origin has remained unexplained. We show below how this spectrum and its $T$ dependence stem from the behavior of the complex phonon self-energy. 

First, we consider the experimental $T$ dependence of TO spectra at zone center. Data for SnTe and PbTe at $\Gamma$-point in multiple Brillouin zones was measured with HB-3, using PG002 monochromator and analyzer, with constant final energy $E_{f}=$ 14.7 meV, and collimation settings of $48'$-$40'$-$40'$-$120'$. The spectra for $\mathbf Q = (1,1,1)$ are shown in Fig.~\ref{fig:TO_113}. Similar spectra were found at other equivalent points (Supplemental Material).

These measured spectra for $ \chi^{\prime \prime}_{{\rm TO}, {\mathbf q}=0}(\it\Omega)$ show the strong anharmonic temperature dependence of the TO peak position and shape, and also reveal important differences between SnTe and PbTe. Overall, the TO spectra are broad, asymmetric (non Lorentzian), and the shift $\it\Delta$ increases with $T$. This is opposite of the prediction from the quasi-harmonic approximation (QHA), which would give ${\it\Delta}/\omega < 0$ and proportional in magnitude to the (positive, increasing) thermal expansion. This deviation from QHA is another manifestation of strong anharmonicity. At high $T$, the spectra for both materials exhibit an asymmetry with a tail of excess intensity on the low-energy side of the peak. On cooling, the peak position decreases in energy as expected, but in addition, the asymmetry reverses, with the tail on the high-energy side (particularly clearly seen at $100\,$K). Also, we re-emphasize that at $300\,$K, the measured TO spectrum in PbTe develops a clear splitting\cite{Delaire:2011gr, Jensen:2012dz}, while there is only a low-energy tail in the case of SnTe, as can be seen in Fig.~\ref{fig:TO_113}-a,b and in Fig.~\ref{fig:SQE_all}-b,e. This effect was reproduced in our first-principles simulations of anharmonicity (Fig.~\ref{fig:TO_113}-c,d and Fig.~\ref{fig:SQE_all}-c,f).

Finite-temperature dynamics were studied computationally with ab-initio Born-Oppenheimer molecular dynamics (MD). Anharmonic terms in interatomic force-constants were obtained by fitting an effective anharmonic hamiltonian to the forces in a set of atomic configurations sampled from the MD trajectories, following the temperature-dependent effective potential (TDEP) methodology developed by Hellman \textit{et al}.\cite{Hellman:PRB2013} Details are presented in Supplemental Material. The anharmonic hamiltonian is then used to obtain renormalized phonon dispersions, as well as the phonon self-energy functions at any $\mathbf Q= \mathbf q + {\mathbf \tau}$ (with ${\mathbf \tau}$ a reciprocal lattice vector).

The INS spectrum of phonon mode $(\mathbf q, j)$ is obtained from the anharmonic self-energy ${\it\Sigma}_{\mathbf q j} = {\it\Delta}_{\mathbf q j} + i {\it\Gamma}_{\mathbf q j}$ \cite{Cowley:RepProgPhys1968}:
\begin{multline}
\chi^{\prime \prime}_{\mathbf q j}( {\mathit\Omega}) = |F(\mathbf Q, {\mathit\Omega})|^2 \times \\
 {\frac{2 \omega_{\mathbf q j} {\it\Gamma}_{\mathbf q j}( \it\Omega)} {{\{\mathit\Omega}^2 - \omega_{\mathbf q j}^2 - 2 \omega_{\mathbf q j} {\mathit\Delta}_{\mathbf q j}({\it\Omega}) \}^2 \, + \, 4 \omega_{\mathbf q j}^2 {\mathit\Gamma}_{\mathbf q j}( {\mathit\Omega})^2} \; ,}
\label{anh-cross-section}
\end{multline}
\noindent where $\omega_{\mathbf q j}$ is the dispersion frequency, ${\mathit \Omega}=E/\hbar$ is the driving frequency with which the system is probed (neutron energy transfer). Because $\omega_{\mathbf q j}$ of the TO mode is imaginary (unstable) in the harmonic approximation, we use instead the stable effective harmonic frequency obtained from TDEP. The structure factor, $F(\mathbf Q, \it\Omega)$ for $\mathbf Q = \mathbf q + \mathbf{\tau}$, was obtained from the corresponding polarizations and energies. The real and imaginary parts of the phonon self-energy, ${\it\Delta}_{\mathbf q j}(\Omega)$ and ${\it\Gamma}_{\mathbf q j}(\Omega)$, are evaluated, to lowest order in perturbation theory, following \cite{Maradudin-Fein:1962, Maradudin-Fein-Vineyard:1962, Cowley:RepProgPhys1968}. For the imaginary part: 
\begin{eqnarray}
\label{eq:gamma}
{{\mathit\Gamma_{\mathbf q j}(\it\Omega)}} &=& {\frac{18}{\hbar^2} \sum_{\mathbf q_1 \mathbf q_2 j_1 j_2} \left| V_3 \left( ^{{\mathbf q} \, {\mathbf q_1} \, {\mathbf q_2}}_{j \, j_1 \, j_2} \right) \right|^2 \times } \\
&& \nonumber { \Big( (n_1 + n_2 +1 ) \left[ \delta_{\omega_1 + \omega_2 - \mathit\Omega} - \delta_{\omega_1 + \omega_2 + \mathit\Omega} \right] } \\
&& \nonumber {+ (n_2 - n_1) \left[ \delta_{\omega_1 - \omega_2 - \mathit\Omega} - \delta_{\omega_1 - \omega_2 + \mathit\Omega} \right] \Big)\; ,}
\end{eqnarray}
\noindent where $V_3$ is the Fourier-transform of the third--order component of the interatomic potential, and the $(\mathbf q_k \, j_k)$ are the different interacting phonon modes, with occupation functions $n_k$. From $\mathit\Gamma_{\mathbf q j}$, we obtained $\mathit\Delta_{\mathbf q j}$ by Kramers--Kronig (Hilbert) transformation, $\mathit\Delta_{\mathbf q j}(\Omega)=\mathcal{H}[\mathit\Gamma_{\mathbf q j}](\Omega)$. The resulting spectra, $\chi^{\prime \prime}_{\mathbf q j}(\Omega)$, are shown in Fig.~\ref{fig:SQE_all}-c,f and Fig.~\ref{fig:TO_113}-c,d. They are in good agreement with the neutron measurement, predicting not only the shift of the peak with temperature, but also the splitting of the spectrum in PbTe. In the following, we leave indices $(\mathbf q, j)$ implicit when considering the TO mode at $\Gamma$.

 \begin{figure}
 \noindent\makebox[\columnwidth]{
\includegraphics[width=1.2\columnwidth]{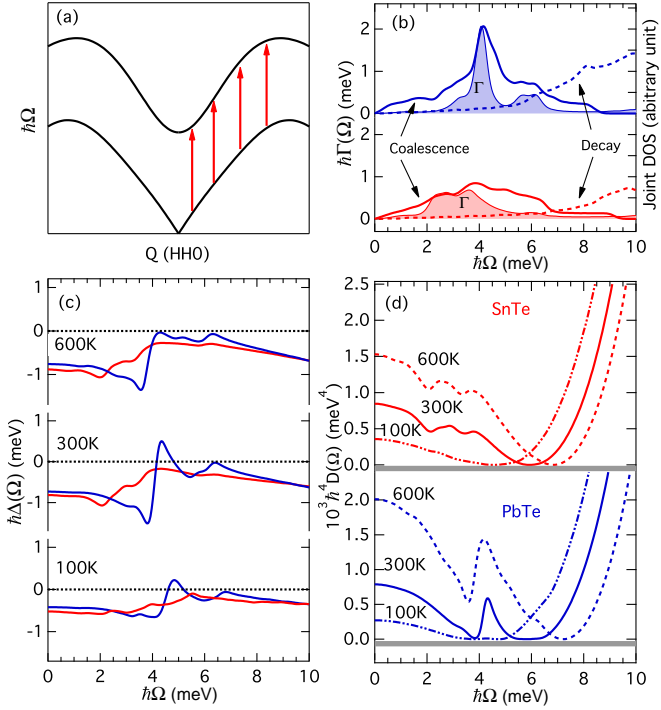}}
\caption{\label{fig:SE} Relation between the phonon self-energy and the complex features in INS spectra. (a) Schematic of nesting between TA and TO branches in phonon dispersions of PbTe. (b) Imaginary part of the phonon self-energy, $\mathit\Gamma$ (shaded area), and joint density of states (lines). Solid (dashed) lines are coalescence (decay) contributions to the joint--DOS. (c) Real part of the phonon self-energy, $\mathit\Delta$. Light red lines are for SnTe and dark blue lines for PbTe. (d) Minima in $D(\it\Omega)$ (Eq. \ref{eq:d}) determine the position of peaks in INS spectra. }
\end{figure}

As can be seen in Fig.~\ref{fig:SE}-b, a sharp peak occurs around 4$\,$meV in $\mathit\Gamma(\Omega)$ for PbTe ($T=300\,$K), while a broader peak is found at 3$\,$meV for SnTe. This peak originates from the joint DOS for ``coalescence'' processes (of the type $\mathit\Omega + \omega_1 \rightarrow \omega_2$), also shown in Fig.~\ref{fig:SE}-b. The joint DOS is proportional to Eq.~\ref{eq:gamma} when setting $V_3$ to a constant, and represents conservation of phonon energy and momentum (it is a measure of the phase-space volume for available phonon-scattering channels). Thus, the sharp peak in $\mathit\Gamma_{\mathbf q=0}(\Omega)$ and the joint DOS reflects the nesting (parallel sections) in phonon dispersions of PbTe (illustrated in Fig.~\ref{fig:SE}-a), with an energy difference $\hbar\mathit\Omega_0 \sim 4\,$meV. The key difference between the two materials is that this peak is sharper in PbTe than in SnTe, because of better nesting of dispersions in PbTe. This peak in $\mathit\Gamma(\Omega)$ leads to a strong oscillation of $\mathit\Delta(\Omega)$. Since the Hilbert transform verifies $\mathcal{H}[\delta(x)]=1/x$, the sharp peak in $\mathit\Gamma$ near $\mathit\Omega_{0}$, leads to $\mathit\Delta \sim1/(\mathit\Omega-\mathit\Omega_{0})$. This strong variation in $\mathit\Delta$, shown in Fig.~\ref{fig:SE}-c, produces the complex shape of the TO spectra, as explained below.

From Eq.~\ref{anh-cross-section}, $\chi^{\prime \prime}(\it\Omega)$ shows a peak where its denominator reaches a minimum. Thus, we can expect a peak in the phonon spectral function where 
\begin{eqnarray}
\label{eq:d}
D({\it\Omega})= \{{ {\it\Omega}^2 - \omega^2 - 2 \omega \it\Delta(\Omega)}\}^2
\end{eqnarray}
has a minimum. This function is shown for different temperatures in Fig.~\ref{fig:SE}-d. The large oscillation in $\it\Delta$ for PbTe at 300$\,$K gives rise to two well-separated minima in $D$, as see in Fig.\ref{fig:SE}-d, and thus to two peaks in $\chi^{\prime \prime}(\it\Omega)$, as observed experimentally. In the case of SnTe, $D$ shows a single minimum, but remains lower on the low-energy side, causing the observed asymmetry in the spectrum $\chi^{\prime \prime}(\it\Omega)$. We emphasize that for the strongly anharmonic TO mode, it is no longer possible to properly describe the phonon spectrum with just a frequency and a single linewidth. Instead, the full functional dependence of $2{\it\Gamma}_{\mathbf q j}(\Omega)$ and ${\it\Delta}_{\mathbf q j}(\Omega)$ are needed to capture the features in INS phonon spectra, as was done here.

We note that the joint DOS at ${\Gamma}$ is proportional to $n(T)$, and thus linear in $T$ at high temperature. On the other hand, because $\omega_{\rm TO, \Gamma}$ increases with $T$, we obtain that $|V_3|^2$ decreases with $T$. As a result, the width of the zone-center TO spectrum (shown as a figure in Supplemental Material) is more weakly $T$-dependent than the linear dependence expected from the Bose distributions in Eq.~\ref{eq:gamma}, as was observed experimentally.\cite{Delaire:2011gr}

In conclusion, the TO phonon spectral function at $\Gamma$-point in SnTe and PbTe were explained by first-principles calculations of the anharmonic phonon self-energy. Kinematics of phonon scattering processes dominate the shape of the self-energy and provided the key to rationalize the anomalies observed experimentally. We emphasize that, while the proximity to the ferroelectric transition is needed to induce anharmonic potentials, the occurrence of the soft-mode is not enough by itself to account for the complex phonon spectra at the zone center. The nesting between phonon dispersions is key to enabling a large number of 3-phonon scattering channels for the TO mode. This effect is stronger in PbTe than SnTe, even though SnTe is closer to the ferroelectric instability, as shown by both experiments and calculations (Supplemental Material). This mechanism is general, and it could be possible to design materials with desired anharmonicity by achieving such nesting of phonon dispersions through electron doping, chemical pressure, and tailoring mass ratios. In addition, first-principles computations of phonon dispersions could systematically screen candidate materials for such occurrences.

\begin{acknowledgments}
This work was supported by the U.S. Department of Energy, Office of Basic Energy Sciences, Materials Sciences and Engineering Division (theory and material synthesis) and the S3TEC Energy Frontier Research Center (neutron measurements). Modeling of neutron data was developed as part of CAMM, funded by the U.S. Department of Energy, Basic Energy Sciences, Materials Sciences and Engineering Division. The research at Oak Ridge National Laboratory's Spallation Neutron Source and High Flux Isotope Reactor was sponsored by the Scientific User Facilities Division, Office of Basic Energy Sciences, U.S. DOE. 
\end{acknowledgments}

\bibliography{snte}

\end{document}